\documentclass[aps,prl,twocolumn,amsmath,amssymb]{revtex4-2}

\usepackage{graphicx}
\usepackage{bm}

\newcommand{\Agw}{A_{\rm gwb}}
\newcommand{\ggw}{\gamma_{\rm gw}}
\newcommand{\logA}{\log_{10}A}
\newcommand{\lnL}{\ln\mathcal{L}}
\newcommand{\lnV}{\ln V}
\newcommand{\Np}{N_p}
\newcommand{\Nf}{N_f}

\begin{document}

\title{Prior volume effect inflates the gravitational-wave background in pulsar timing arrays}

\author{Biwei Dai}
\email{biwei@ias.edu}
\author{Matias Zaldarriaga}
\affiliation{%
 School of Natural Sciences, Institute for Advanced Study, 1
Einstein Drive, Princeton, NJ 08540, USA
}%
\author{Yacine Ali-Ha\"{\i}moud}
\affiliation{%
 Center for Cosmology and Particle Physics, Department of Physics, New York University, New York, NY 10003, USA}

\date{\today}

\begin{abstract}

Pulsar timing arrays (PTAs) have found evidence for a nanohertz gravitational-wave background (GWB) whose amplitude exceeds most predictions based on supermassive black hole binaries (SMBHBs).
Using NANOGrav 15-year data, we show that a substantial part of this excess arises from a prior volume effect. The GWB amplitude is constrained mainly by the pulsars' auto-correlations, where the GWB is degenerate with each pulsar's intrinsic red noise. By decomposing the GWB posterior, we find that marginalizing over the standard log-uniform red-noise priors introduces a prior volume term that favors a stronger GWB, pulling the inferred amplitude more than $2\sigma$ above the maximum-likelihood estimate. The Hellings-Downs (HD) cross-correlations, which measure the GWB alone, are not strong enough to counter the volume term. The profile likelihood, a cross-correlation-only estimator, and priors that are uniform in power or informed by the pulsar population all avoid or reduce this effect and lower the amplitude into agreement with SMBHBs. The revised priors also yield a steeper GWB spectrum closer to the SMBHB predictions, strengthen the evidence for HD correlations, and weaken the preference for new-physics sources over SMBHBs.

\end{abstract}

\maketitle

\section{\label{intro}Introduction}

Multiple pulsar timing array (PTA) collaborations have reported evidence for a gravitational wave background (GWB) at nanohertz frequencies~\cite{NANOGrav15_gwb,EPTA_gwb,PPTA_gwb,CPTA_gwb}. The signal is broadly consistent with a cosmic population of supermassive black hole binaries (SMBHBs), but its inferred amplitude sits at or above the upper end of most population-synthesis predictions~\cite{Sesana2013,Shannon2015,Simon2016,Middleton2021}, in tension with the observed SMBH population at the $2$--$4.5\sigma$ level~\cite{Sato-Polito2024}. This excess has motivated explanations invoking additional GWB sources from the early universe such as cosmic strings and phase transitions~\cite{NANOGrav15_newphysics,EPTA_interpretation}.

The GWB amplitude is inferred jointly with each pulsar's intrinsic red noise (RN). The standard analysis models every pulsar's RN as a power law and marginalizes over its amplitude under wide, mutually independent log-uniform priors~\cite{NANOGrav15_gwb}. 
This choice was motivated by early simulation studies with little or no intrinsic RN injected~\cite{vanHaasteren2013,Hazboun2020}, for which a prior favoring small RN amplitudes like log-uniform is appropriate.
However, in practice most pulsars show significant RN~\cite{NANOGrav15_noise}, and their RN parameters are drawn from a common Galactic population that is unlikely to be log-uniform. This has motivated hierarchical and population-level treatments of PTA noise~\cite{vanHaasteren2024,Goncharov2024,Goncharov2022,DAmico2025,Villa2025,Goncharov2025}. Here we show that marginalizing over these broad log-uniform priors, which do not describe this population, introduces a prior volume term that inflates the GWB amplitude. 

Prior volume effects, also called projection effects, are well known in cosmological data analysis~\cite{Handley2019,Hergt2021,Simon2023,Hadzhiyska2023,Holm2023}, where marginalizing over poorly-constrained nuisance parameters shifts posteriors away from the likelihood peak. Here we present the first explicit decomposition of the NANOGrav posterior into profile-likelihood and volume contributions, and explain why the volume term favors a stronger GWB. We then mitigate the effect along four different routes. All these reanalyses recover a lower GWB amplitude consistent with SMBHB predictions. The revised priors also strengthen the evidence for Hellings-Downs (HD) correlations and weaken the preference for new-physics GWB sources. 

\section{Prior volume effect in PTA}\label{sec:decomposition}

Consider a PTA model with GWB parameters $\bm{\theta} = (\logA_{\rm gwb}, \ggw)$ and per-pulsar RN parameters $\bm{\phi} = \{(\logA_i, \gamma_i)\}_{i=1}^{\Np}$. The data $d$ are the pulsars' timing residuals, described by a Gaussian likelihood $\mathcal{L}(d|\bm\theta,\bm\phi)$ marginalized over each pulsar's deterministic timing model~\cite{Lentati2013,vanHaasterenVallisneri2014,NANOGrav15_gwb,NANOGrav15_noise}. Each pulsar's intrinsic RN and the common GWB are modeled as stationary Gaussian processes with power-law timing-residual power spectral densities,
\begin{equation}\label{eq:powerlaw}
    P(f) = \frac{A^2}{12\pi^2}\left(\frac{f}{f_{\rm yr}}\right)^{-\gamma} f_{\rm yr}^{-3}\,,
\end{equation}
where $f_{\rm yr}\equiv1\,{\rm yr}^{-1}$, and amplitude $A$ and slope $\gamma$ are the free parameters. For the GWB the inter-pulsar cross-power additionally carries the Hellings-Downs angular correlation~\cite{HellingsDowns1983}, and a slope $\ggw=13/3$ corresponds to a population of circular, GW-driven SMBHB inspirals~\cite{Phinney2001}. The marginal posterior of GWB parameters is
\begin{equation}\label{eq:posterior}
    p(\bm{\theta} | d) \propto \pi_{\rm gwb}(\bm{\theta}) \int \mathcal{L}(d | \bm{\theta}, \bm{\phi})\, \pi_{\rm rn}(\bm{\phi})\, d\bm{\phi}\,,
\end{equation}
where $\pi_{\rm gwb}$ and $\pi_{\rm rn}$ are the GWB and RN priors, both uniform in $\log_{10}A$ and $\gamma$ in the standard analysis, with $\pi_{\rm rn}$ independent across pulsars.

\subsection{Marginal posterior decomposition}

\begin{figure*}[t]
    \centering
    \includegraphics[width=\textwidth]{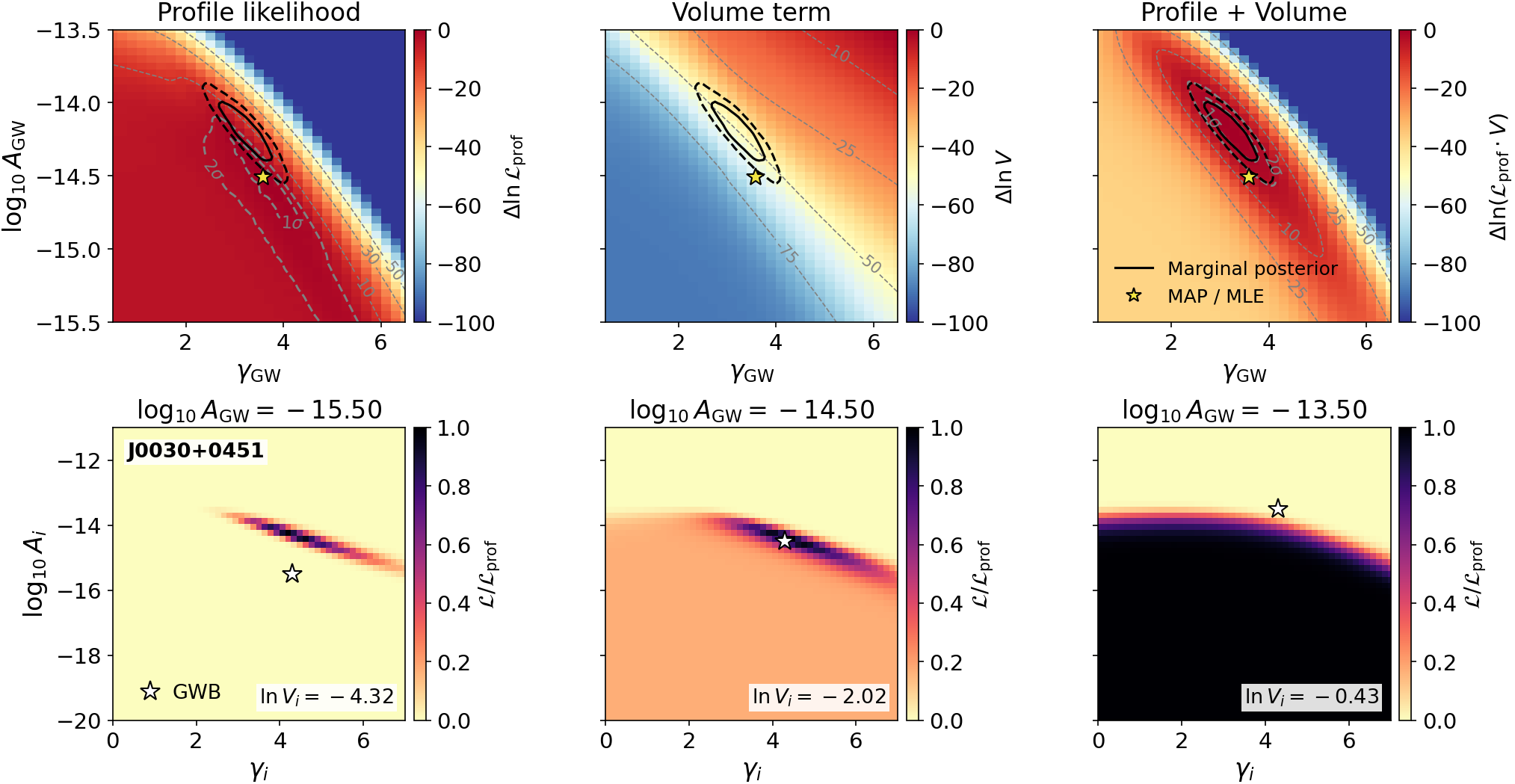}
    \caption{\textbf{Top:} decomposition of the HD marginal posterior on NANOGrav 15-year data into the profile likelihood (left), the volume term (center), and their sum (right), each normalized to its maximum. Grey $1\sigma$ and $2\sigma$ contours in the left panel denote the profile-likelihood constraints. The volume increases monotonically towards larger GWB, pulling the marginal posterior (black contour) up by more than $2\sigma$ compared to the MAP/MLE (star). \textbf{Bottom:} the Eq.~\ref{eq:Vi} volume integrand ($\mathcal{L}(d|\bm\theta,\phi_i,\{\hat\phi_j\})/\mathcal{L}_{\rm prof}(\bm\theta)$) for pulsar J0030+0451 over its RN parameters, at three representative GWB levels (indicated by stars). The allowed RN space grows with GWB, explaining why the volume term prefers higher GWB.}
    \label{fig:decomposition}
\end{figure*}

The marginal posterior in Eq.~(\ref{eq:posterior}) can be decomposed as~\cite{Hadzhiyska2023}
\begin{equation}\label{eq:decomposition}
    \ln p(\bm{\theta} | d) = \lnL_{\rm prof}(\bm{\theta}) + \lnV(\bm{\theta}) + \ln\pi_{\rm gwb}(\bm{\theta}) + \text{const}\,,
\end{equation}
where $\mathcal{L}_{\rm prof}(\bm{\theta}) \equiv \mathcal{L}(d | \bm{\theta}, \hat{\bm{\phi}}(\bm{\theta}))$ is the profile likelihood with $\hat{\bm{\phi}}(\bm{\theta}) = \arg\max_{\bm{\phi}}\, \mathcal{L}(d | \bm{\theta}, \bm{\phi})$, and the volume term is 
\begin{equation}\label{eq:volume}
    V(\bm{\theta}) = \int \frac{\mathcal{L}(d|\bm{\theta},\bm{\phi})}{\mathcal{L}_{\rm prof}(\bm{\theta})}\, \pi_{\rm rn}(\bm{\phi})\, d\bm{\phi}\,.
\end{equation}
The profile likelihood captures ``what the data prefer'' given the best possible nuisance parameters, while $V(\bm\theta)$ captures ``how much prior volume is available'' in the nuisance RN parameter space. Note that $V(\bm\theta)$ depends on the RN prior, while $\mathcal{L}_{\rm prof}(\bm{\theta})$ does not.

$V(\bm\theta)$ is a high-dimensional integral, which we factorize across pulsars,
\begin{equation}\label{eq:volume_factorize}
    \lnV(\bm{\theta}) \approx \sum_{i=1}^{\Np} \lnV_i(\bm{\theta})\,,
\end{equation}
where each pulsar contributes
\begin{equation}\label{eq:Vi}
    V_i(\bm{\theta}) = \int \frac{\mathcal{L}\big(d\,|\,\bm{\theta},\,\phi_i,\,\{\hat\phi_j\}_{j\neq i}\big)}{\mathcal{L}_{\rm prof}(\bm{\theta})}\, \pi_{\rm rn}(\phi_i)\, d\phi_i\,.
\end{equation}
Here the numerator is the \emph{full} PTA likelihood with only pulsar $i$'s two RN parameters $\phi_i$ varying, while every other pulsar is pinned at its profile-maximum value $\hat\phi_j(\bm{\theta})$. This factorization is exact when, at fixed $\bm\theta$, different pulsar's RN parameters are independent, as in the common uncorrelated red noise (CURN) model. For HD, marginalizing over GW realizations could potentially couple different pulsars' RN parameters, but we verify from our chains that these correlations are negligible, so it remains a good approximation. Note that we neglect only those RN-parameter correlations, not the HD correlations themselves, which enter through the full-PTA likelihood in Eq.~(\ref{eq:Vi}).

We apply this decomposition to the NANOGrav 15-year public dataset~\cite{NANOGrav15_data}, following the modeling pipeline of the official analysis~\cite{Johnson2024}. The top row of Fig.~\ref{fig:decomposition} shows the result in the $(\ggw, \logA_{\rm gwb})$ plane. The profile likelihood (left) is flat where the GWB is too weak to matter, peaks around the maximum-likelihood estimate ($\logA_{\rm gwb} = -14.51$, $\ggw = 3.6$, yellow star), and cuts off where the GWB would overpredict the data. The volume term (center) instead grows monotonically toward stronger GWB, pulling the posterior peak above the MLE by more than $2\sigma$. Their sum (right) reproduces the MCMC marginal posterior and validates the decomposition and pulsar factorization.

\subsection{Why the volume favors a stronger GWB}\label{sec:origin}

The bottom row of Fig.~\ref{fig:decomposition} shows the volume integrand (Eq.~\ref{eq:Vi}) for one pulsar at three GWB amplitudes. Within a pulsar the intrinsic RN and the GWB both contribute red power, and the data constrain only the sum $\Agw^2+A_{\rm rn}^2$. When the GWB is weak (left), the RN is pinned by the observed red power and the volume is small. As the GWB supplies a growing share (center), smaller RN amplitudes become allowed, and once it accounts for all the observed power (right) the entire low-amplitude range of the RN prior contributes. This picture explains why volume grows with $\Agw$, and since the total volume is the product of the per-pulsar factors (Eq.~\ref{eq:volume_factorize}), the effect compounds over the $\Np$ pulsars (per-pulsar breakdown in Fig.~\ref{fig:volume}).

\emph{Analytical model.}---The same mechanism can be understood in a simplified model (End Matter). Keep only the GWB-dominated low-frequency band and describe each process by a single power, $\Agw^2$ for the GWB and $A_{{\rm rn},i}^2$ for pulsar $i$'s intrinsic RN. The auto-correlations then measure only each pulsar's total power, $A_{{\rm tot},i}^2 \equiv \Agw^2 + A_{{\rm rn},i}^2$, while the Hellings-Downs cross-correlations measure $\Agw^2$ alone. We assume, as in current PTAs and in the derivation of the optimal statistic~\cite{Anholm2009,Chamberlin2015}, that the cross-correlations carry much less information than the auto-correlations. The likelihood then factorizes into per-pulsar auto-correlation terms $L^i_{\rm auto}(A_{{\rm tot},i}^2)$ and a weak cross-correlation likelihood $L_{\rm cross}(\Agw^2)$ (End Matter),
\begin{equation}\label{eq:model_likelihood}
    p(d|\Agw^2,\{A_{{\rm rn},i}^2\}) \approx L_{\rm cross}(\Agw^2) \prod_{i} L^i_{\rm auto}(A_{{\rm tot},i}^2)\,.
\end{equation}
Marginalizing over the RN powers with the RN prior $\pi_{\rm rn}$ written as a density $\pi_{A^2}$ in the power leads to the marginal posterior
\begin{equation}\label{eq:ridge}
\begin{aligned}
    p(\Agw^2 | d) &\propto \pi_{\rm gwb}(\Agw^2)\, L_{\rm cross}(\Agw^2) \prod_{i=1}^{\Np} P^i_{\rm auto}(\Agw^2)\,,\\
    P^i_{\rm auto}(\Agw^2) &= \int\! dA_{{\rm rn},i}^2\; L^i_{\rm auto}(\Agw^2 + A_{{\rm rn},i}^2)\,\pi_{A^2}(A_{{\rm rn},i}^2)\,.
\end{aligned}
\end{equation}
For simplicity, we take the many-mode limit and the auto-correlations pin $A_{{\rm tot},i}^2 = A_{{\rm rn},i}^2 + \Agw^2$ ($L^i_{\rm auto}$ becomes a delta function) and $P^i_{\rm auto}(\Agw^2) \to \pi_{A^2}(A_{{\rm tot},i}^2 - \Agw^2)$. Note that our conclusion does not rely on this limit, as is shown in the End Matter. For a power-law RN prior $\pi(A) \propto A^{p}$, i.e.\ $\pi_{A^2}(A^2) \propto (A^2)^{(p-1)/2}$, we have 
\begin{equation}\label{eq:ridge2}
    P^i_{\rm auto}(\Agw^2) \propto (A_{{\rm tot},i}^2 - \Agw^2)^{-(1-p)/2}
\end{equation}
for $\Agw^2 \le \min_i A_{{\rm tot},i}^2$.
 
\begin{figure}[t]
    \centering
    \includegraphics[width=\columnwidth]{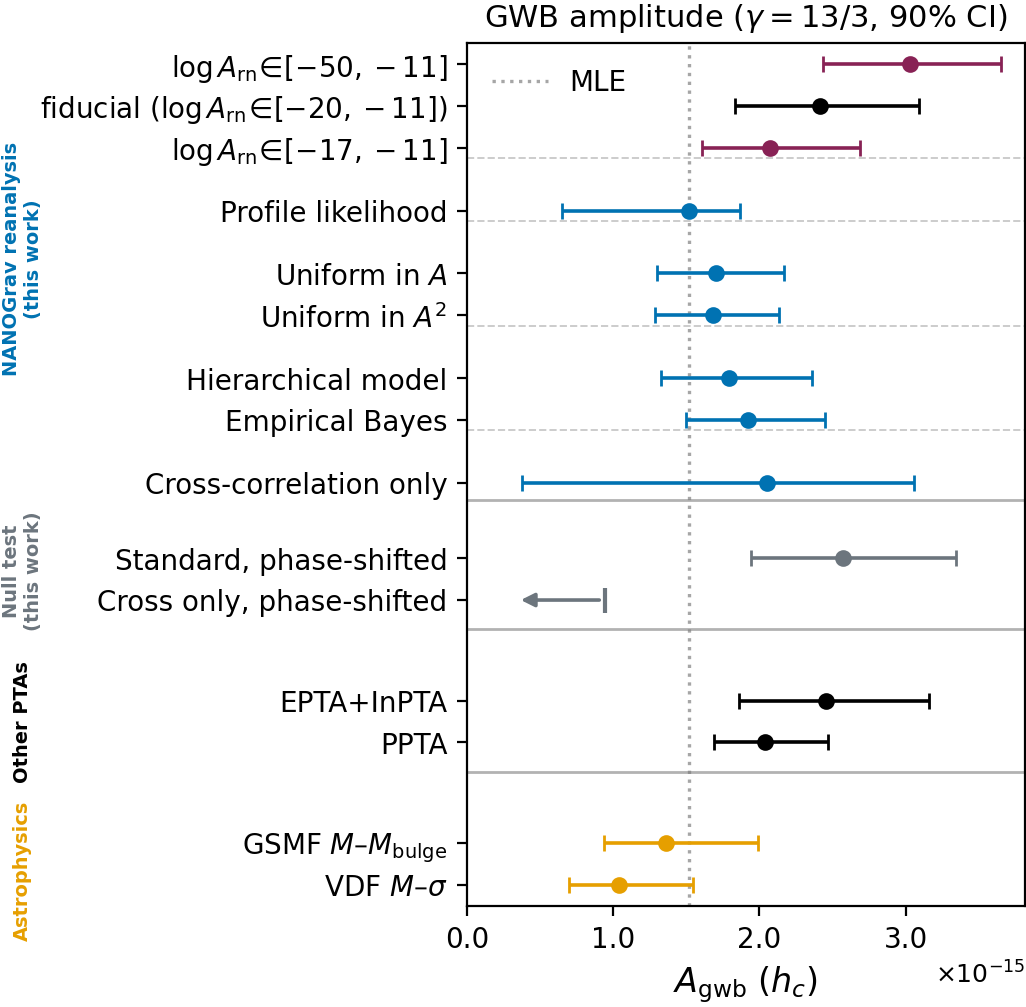}
    \caption{GWB amplitude $\Agw$ constraints with $\ggw = 13/3$, as 90\% credible or confidence intervals. The top three rows use the standard log-uniform RN prior with different lower bounds, which change the prior volume and shift the GWB constraint. The fiducial row is the official NANOGrav choice. The mitigation methods (blue) all yield a lower inferred GWB amplitude, consistent with the predictions from galaxy scaling relations (orange)~\cite{Sato-Polito2024}. The null test (grey) repeats the standard and cross-correlation-only analyses after removing the inter-pulsar correlations. The standard amplitude barely changes, while the cross-correlation amplitude is consistent with zero and shown as a 95\% upper limit. EPTA+InPTA~\cite{EPTA_gwb} and PPTA~\cite{PPTA_gwb} use standard log-uniform priors.}
    \label{fig:fixgamma}
\end{figure}

Combining Eq.~\ref{eq:ridge} and \ref{eq:ridge2}, we see that any prior with $p<1$ tilts the posterior toward large $\Agw^2$ compared to $L_{\rm cross}(\Agw^2)$, and because each pulsar contributes an independent factor the tilt compounds over the array. For the standard log-uniform prior with $p=-1$, the prior mass around the GWB upper bound $\Agw^2 \sim \min_i A_{{\rm tot},i}^2$ depends on the RN prior lower bound $A_{{\rm rn},\min}^2$ (which can also be seen in Fig.~\ref{fig:decomposition}). We show this explicitly on the real data: widening the lower bound of the RN amplitude prior pushes the posterior to higher $\Agw$, while truncating it lowers the inferred amplitude (top three rows of Fig.~\ref{fig:fixgamma}), even though both bounds are at least two orders of magnitude smaller than the signal and too small to change the likelihood meaningfully.

This strong-GWB preference would be correct if each pulsar's RN were truly log-uniform down to arbitrarily low amplitudes, since a common red-power floor is then more economically explained by a shared GWB than by a coincidence of independent RN amplitudes, in the spirit of Occam's razor. But the pulsar population may itself have a low-amplitude floor, in which case that floor reflects population structure rather than a GWB. The log-uniform prior assumes the former and amplifies the preference beyond what the population warrants.

One might hope the HD cross-correlations break the degeneracy between the GWB and the per-pulsar RN. To test this on the data, we apply an independent random phase shift to each pulsar's GWB Fourier basis~\cite{Taylor2017},
which destroys the inter-pulsar phase coherence that HD correlations require while preserving the auto-power, and rerun the HD inference. Removing the cross-correlations barely changes the inferred GWB when using the standard priors (first gray interval in Fig.~\ref{fig:fixgamma}), and both agree with the CURN analysis. The amplitude is thus set almost entirely by the auto-correlated power, the channel degenerate with per-pulsar RN~\cite{Romano2021}. This also verifies our assumption about the cross-correlation being much weaker than the auto-correlation in the analytical model. The cross-correlations give strong evidence that the HD signal exists \cite{NANOGrav15_gwb} but are not enough to overcome the prior volume effect in parameter inference.

\section{Mitigating the prior volume effect}\label{sec:mitigation}

We now introduce the four approaches we used in this work to mitigate the prior volume effect.

\emph{Profile likelihood.}---The profile likelihood (defined under Eq.~\ref{eq:decomposition}) is a frequentist approach that has no prior dependence, and is widely adopted in cosmology for this reason~\cite{Planck2013_profile,Herold2022,GomezValent2022,Holm2023,Herold2025}. We convert it to confidence intervals with Wilks' theorem~\cite{Wilks1938} (Supplemental Material).

\emph{Power-space priors.}---The PTA data directly constrain the power spectrum: the amplitude $A$ enters the likelihood only through the timing-residual spectrum (Eq.~\ref{eq:powerlaw}), which is linear in $A^2$. A prior flat in the GWB and RN power $A^2$ is therefore a natural uninformative choice. This can be verified in the analytical model of Eq.~(\ref{eq:ridge}), where a uniform in $A^2$ prior ($p=1$) does not introduce any tilt.

\emph{Population priors.}---%
The pulsar noise properties are expected to be drawn from a population with characteristic scales~\cite{Goncharov2024,Chalumeau2022}. In this work we implement a population model in two ways. The hierarchical model~\cite{ThraneTalbot2019,vanHaasteren2024,DAmico2025,Villa2025} takes each $(\gamma_i, \logA_i)$ to be drawn from a bivariate Gaussian whose population parameters are sampled jointly with the GWB and marginalized in the inference. 
The empirical Bayes (EB) approach instead estimates the population density non-parametrically as the density $\hat p$ maximizing the product of per-pulsar evidences $\prod_k \int \mathcal{L}_k(d_k|\bm\phi)\,p(\bm\phi)\,d\bm\phi$, obtained from an expectation-maximization iteration~\cite{Laird1978}  and then adopted as a fixed prior. 
Unlike the hierarchical model it can represent a multi-modal population with a 2D histogram, but it ignores the uncertainty in the estimated density.
In the Supplemental Material we describe the iteration and prove that the EB density converges to the true population distribution as $\Np \to \infty$.

\emph{Decoupled HD model.}---We separate the auto- and cross-correlation amplitudes $A_{\rm auto}$ and $A_{\rm cross}$~\cite{NANOGrav12_gwb} and sample them independently, so $A_{\rm cross}$ is inferred from the cross-correlations alone, free of the auto-power volume effect:
\begin{equation}\label{eq:cross_only}
    C_{ab}(f) = \left(A_{\rm auto}^2\, \delta_{ab} + A_{\rm cross}^2\, \Gamma_{\rm HD}(\theta_{ab})\,(1-\delta_{ab})\right)S(f)\,,
\end{equation}
where $S(f)$ is the power law frequency dependence.
A null test in Fig.~\ref{fig:fixgamma} confirms the separation: when we apply the same random phase shifts to destroy the cross-correlations, $A_{\rm cross}$ becomes consistent with $0$. The $A_{\rm auto}$ constraint agrees with the standard analysis to within a few percent, with or without the phase shift, further evidence that the standard analysis is dominated by the auto-correlations. The price of separation is precision, since the cross channel carries far less information than the auto-power. The decoupled model is the Bayesian counterpart of the optimal statistic~\cite{Anholm2009,Chamberlin2015,Vigeland2018,Vallisneri2023,Sardesai2023}, the frequentist cross-correlation estimator, and shares its immunity to the auto-power prior.

\begin{figure}[t]
    \centering
    \includegraphics[width=\columnwidth]{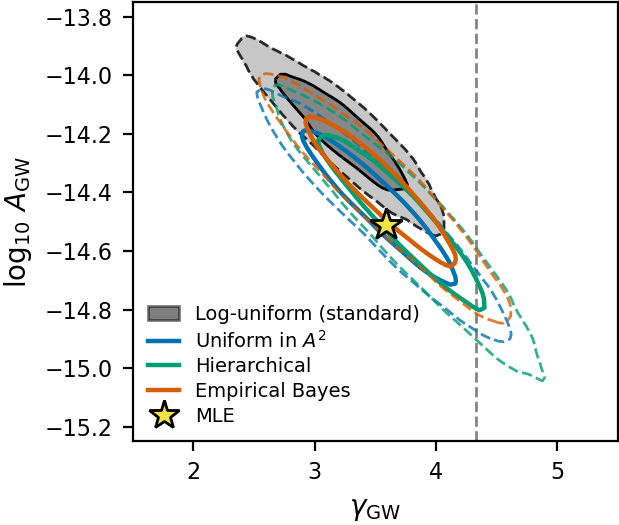}
    \caption{The GWB posterior on NANOGrav 15yr data under different priors. The vertical dashed line marks $\ggw = 13/3$, the value expected for a population of circular, GW-driven SMBHB inspirals. Every revised prior shifts the posterior to a lower amplitude and larger $\ggw$ closer to $13/3$.}
    \label{fig:prior_comparison}
\end{figure}

\emph{GWB inference.}---Figure~\ref{fig:fixgamma} shows the $A_{\rm gwb}$ 90\% credible interval with $\ggw = 13/3$. The standard log-uniform prior gives $2.4^{+0.7}_{-0.6}\times10^{-15}$, comparable to EPTA+InPTA and PPTA under the same prior. Every approach introduced above lowers the amplitude by 1--2$\sigma$, consistent with each other and with the MLE: $1.7^{+0.5}_{-0.4}$ for uniform in $A^2$, $1.8^{+0.6}_{-0.5}$ for the hierarchical model, $1.9^{+0.5}_{-0.4}$ for empirical Bayes, and $A_{\rm cross}=2.1^{+1.0}_{-1.7}$ for the decoupled model, all in units of $10^{-15}$, against the prior-free profile likelihood at $1.5^{+0.3}_{-0.9}\times10^{-15}$. The uniform-in-$A$ prior gives similar constraints, though it is less motivated. Together these overlap the predictions from galaxy scaling relations~\cite{Sato-Polito2024} (the uncertainties include the spread between scaling relations, see Supplemental Material), which are in $2.0\sigma$ and $2.9\sigma$ tension with the fiducial analysis.
With $\ggw$ free (Fig.~\ref{fig:prior_comparison}), each prior moves the posterior toward a smaller amplitude and a steeper slope near $\ggw=13/3$, so the effect distorts the whole spectrum and not the amplitude alone.

\emph{Evidence for Hellings-Downs correlations.}---A lower amplitude might be expected to weaken the evidence for HD correlations. Table~\ref{tab:bf_hdcurn} shows the opposite: the revised priors \emph{raise} the Bayes factor between the HD and CURN models by a factor $1.5$ for uniform in $A^2$ and $1.7$ for the hierarchical model.
The likelihood ratio $\mathcal{L}_{\rm HD}/\mathcal{L}_{\rm CURN}$ is peaked where the GWB amplitude matches the measured cross-correlation amplitude, and falls off when the GWB is either too large or too small. 
Since the evidence is dominated by the parameter region holding most of the posterior mass, the revised priors, which place the posterior near this peak, yield a larger Bayes factor than the standard prior, which pushes it to higher amplitudes where the ratio is smaller (Supplemental Material).
The volume effect thus does more than inflate the amplitude: by displacing the posterior to an amplitude the cross-correlations do not support, it also suppresses part of the evidence that identifies the signal as a GWB. 

\begin{table}[t]
    \centering
    \caption{Bayes factor between the HD and CURN models on NANOGrav 15-year data under different priors, and the ratio to the standard log-uniform prior. Revised priors strengthen the evidence for Hellings-Downs correlations. The standard-prior BF reported here is consistent with the official analysis~\cite{NANOGrav15_gwb} within $1\sigma$. 
    Here we exclude the empirical Bayes prior, since it is estimated assuming the HD correlation which favors the HD model.
    }
    \label{tab:bf_hdcurn}
    \begin{tabular*}{\columnwidth}{@{\extracolsep{\fill}}lcc}
    \hline
    RN prior & $\mathrm{BF}(\mathrm{HD}/\mathrm{CURN})$ & Ratio \\
    \hline
    Standard log-uniform  & $176\pm7$  & $1$ \\
    Uniform in $A^2$      & $263\pm4$  & $1.49\pm0.06$ \\
    Hierarchical Gaussian & $300\pm6$  & $1.70\pm0.08$ \\
    \hline
    \end{tabular*}
\end{table}

\emph{New-physics interpretation.}---The NANOGrav 15-year new-physics search~\cite{NANOGrav15_newphysics} reports Bayes factors favoring several cosmological sources over the SMBHB baseline. Here we test if this preference is driven by the prior volume effect, which raises the inferred GWB and contributes to the tension with SMBHB predictions~\cite{NANOGrav15_astro,holodeck}.
We recompute $\mathrm{BF}$(NewPhysics/SMBHB) with a factorized free-spectrum refit~\cite{Lamb2023} under different RN priors (Table~\ref{tab:newphysics}). 
Compared to the standard log-uniform prior, the revised priors reduce the BF by one to two orders of magnitude, to order unity. The new-physics preference of the standard analysis is therefore largely driven by the same prior volume effect.

\begin{table}[t]
    \centering
    \caption{Bayes factors $\mathrm{BF}(\mathrm{model}/\mathrm{SMBHB})$ for six new-physics GWB sources~\cite{NANOGrav15_newphysics} (the highest-BF variant of each source class) under different RN priors.  The uniform in $A^2$ and population priors bring the BF to $\mathcal{O}(1)$. All fifteen models, including uncertainties (typically $25$--$35\%$), are given in the Supplemental Material.
    }
    \label{tab:newphysics}
    \small
    \begin{tabular*}{\columnwidth}{@{\extracolsep{\fill}}lcccc}
    \hline
     & & Uniform & \multicolumn{2}{c}{Population} \\
    Source & Standard & in $A^2$ & Hier. & EB \\
    \hline
    Scalar-induced       & $57$ & $0.50$ & $2.3$ & $4.6$ \\
    Superstrings         & $32$ & $0.92$ & $4.2$ & $7.0$ \\
    Metastable           & $23$ & $0.79$ & $1.8$ & $3.1$ \\
    Domain walls         & $19$ & $0.36$ & $0.79$ & $1.4$ \\
    Phase transition     & $15$ & $0.31$ & $1.0$ & $1.7$ \\
    Inflation            & $6.4$ & $0.06$ & $0.27$ & $0.5$ \\
    \hline
    \end{tabular*}
\end{table}

\section{Discussion and conclusions}

We have shown that the standard PTA analysis with broad log-uniform RN prior is subject to a prior volume effect. On NANOGrav 15-year data, this effect shifts the GWB amplitude more than $2\sigma$ above the MLE. The per-pulsar RN is degenerate with the GWB in the auto-correlations, and marginalizing over its priors introduces a volume term favoring a stronger GWB (Fig.~\ref{fig:decomposition}) that the HD cross-correlations are too weak to overcome (null test in Fig.~\ref{fig:fixgamma}). This is not a flaw of Bayesian inference but a consequence of the prior choice in high dimensions.

Our four routes to mitigate the prior volume effect converge on a lower amplitude that overlaps the SMBHB predictions (Fig.~\ref{fig:fixgamma}): power-space priors, population priors, the prior-free profile likelihood, and the decoupled HD model, with the last two immune to the prior volume by construction. 
With $\ggw$ free, the revised priors yield a steeper slope, closer to the $13/3$ expected for circular, GW-driven SMBHBs (Fig.~\ref{fig:prior_comparison}).
They also recover part of the HD evidence that the volume effect had suppressed (Table~\ref{tab:bf_hdcurn}) and weaken the preference for new-physics sources (Table~\ref{tab:newphysics}).

Simply adding pulsars will not remove the prior-volume shift, since each new pulsar contributes its own volume term (Eq.~\ref{eq:volume_factorize}). As we show in the End Matter, the volume is dominated by the best-timed pulsars, whose red power is well measured and not much larger than the GWB. Better data and longer baselines would strengthen the cross-correlations but also move more pulsars into this regime, so it is not clear whether they will reduce the volume effect as long as the log-uniform RN prior is used. The approaches proposed in this work provide a principled way to quantify and isolate the prior volume effect. We recommend that future PTA analyses report the profile likelihood alongside the marginal posterior, adopt power-space or population-level RN priors, and use the decoupled HD model to isolate the cross-correlation amplitude.

A recent NANOGrav reanalysis with customized chromatic noise models lowers the fixed-$\ggw$ amplitude to $2.1^{+0.6}_{-0.5}\times10^{-15}$~\cite{NANOGrav15_chromatic}, through a mechanism different from the prior volume effect. On EPTA DR2 data, Goncharov et al.~\cite{Goncharov2025} apply the hierarchical noise framework~\cite{vanHaasteren2024,Goncharov2024} and find the posterior moving along the $\Agw$--$\ggw$ degeneracy toward a smaller amplitude and larger slope, but their fixed-$\ggw$ amplitude stays unchanged, unlike ours (Fig.~\ref{fig:fixgamma}). This may stem from the different treatment of the population hyperparameters and from EPTA having fewer pulsars with free RN. Understanding this difference, applying our framework to other PTA datasets, and combining it with improved chromatic noise models remain for future work. 

\begin{acknowledgments}

We thank Tristan Smith and Gabriela Sato-Polito for discussions on the impact of priors on PTA analysis, Nickolas Kokron for discussions on prior volume effects, Aaron Johnson for help with the NANOGrav analysis, and Rutger van Haasteren, Nianyi Chen, and Sihao Cheng for helpful discussions. B.D. acknowledges support from the Ambrose Monell Foundation, the Corning Glass Works Foundation Fellowship Fund, and the Institute for Advanced Study. M.Z. acknowledges support from the Nelson Center for Collaborative Research at the Institute for Advanced Study and the Simons Foundation through Award No. SFI-MPS-BH-00012593-10. M.Z. acknowledges support from the National Science Foundation NSF-BSF 2207583 and NSF 2209991. We used Claude Code (Anthropic) to write some of the analysis and plotting code and to edit the manuscript text. The authors designed the analyses, directed the tool, reviewed its code, and verified all results and numbers reported here.

\end{acknowledgments}

\bibliography{references}

\clearpage

\appendix

\section{Analytical model of the prior volume effect}\label{app:toy}

Here we provide more details on the analytical model of Eq.~(\ref{eq:model_likelihood}). The simplified model takes the intrinsic RN and the GWB to share one spectral shape, so that each process is described by a single power.

\emph{Single-bin covariance.}---In an $\Np$ pulsar array, at Fourier frequency $f_k$, the $\Np$-vector of Fourier coefficients $d$ is a Gaussian variable drawn from $\mathcal{N}(0, \Sigma_k)$ with
\begin{equation}\label{eq:toy_cov}
\begin{aligned}
    \Sigma_k = s_k\big[\Agw^2\,M &+ \mathrm{diag}(A_{{\rm rn},1}^2, \dots, A_{{\rm rn},\Np}^2)\big] \\
    &+ \mathrm{diag}\big(W_1, \dots, W_{\Np}\big)\,,
\end{aligned}
\end{equation}
where $s_k$ is the common power-law shape at $f_k$, $M$ is the Hellings-Downs matrix with $M_{aa} = 1$ and $M_{ab} = \Gamma_{\rm HD}(\theta_{ab})$ for $a \ne b$, and $W_i$ is pulsar $i$'s white-noise variance. The Gaussian log-likelihood is given by
\begin{equation}\label{eq:toy_logL}
\begin{aligned}
    \ln L_k &= -\tfrac12\sum[\ln|\Sigma_k| + d^{\mathsf T}\Sigma_k^{-1}d] + C \\
    &= -\tfrac12\sum[\ln|\Sigma_k| + \mathrm{Tr}(\Sigma_k^{-1}\mathbf S_k)] + C,
\end{aligned}
\end{equation}
where we have used $d^{\mathsf T}\Sigma_k^{-1}d = \mathrm{Tr}(\Sigma_k^{-1}dd^{\mathsf T})=\mathrm{Tr}(\Sigma_k^{-1}\mathbf S_k)$ and $\mathbf S_k$ is the sample covariance. Thus the data enter only through $\mathbf S_k$, whose diagonal and off-diagonal entries are the measured auto- and cross-powers. 

\emph{Factorization of the likelihood.}---Write $\Sigma_k = D_k + E_k$ with $D_k = \mathrm{diag}\big(s_k A_{{\rm tot},i}^2 + W_i\big)$ and $E_{k,ab} = s_k\Agw^2\,\Gamma_{\rm HD}(\theta_{ab})$ for $a \ne b$, $E_{k,aa} = 0$. In the weak-signal regime, we can expand it to second order in $E_k$ (using $\mathrm{Tr}(D_k^{-1}E_k) = 0$)
\begin{equation}\label{eq:toy_factorize}
\begin{aligned}
    \ln L_k &= \sum_{i=1}^{\Np} \ln L^i_{{\rm auto},k}(A_{{\rm tot},i}^2) + \ln L_{{\rm cross},k}\,,\\
    \ln L^i_{{\rm auto},k} &= -\ln\big(s_k A_{{\rm tot},i}^2 + W_i\big) - \frac{S_{k,ii}}{s_k A_{{\rm tot},i}^2 + W_i}\,,\\
    \ln L_{{\rm cross},k} &= \mathrm{Tr}(D_k^{-1}E_kD_k^{-1}\mathbf S_k) + \tfrac12\mathrm{Tr}\big((D_k^{-1}E_k)^2\big) \\
    &\quad - \mathrm{Tr}(D_k^{-1}E_kD_k^{-1}E_kD_k^{-1}\mathbf S_k) + \mathcal{O}(E_k^3)\,,
\end{aligned}
\end{equation}
which gives Eq.~(\ref{eq:model_likelihood}).
The zeroth-order term factorizes exactly over pulsars, each factor depending on the pulsar's total red power alone. With $S_{k,ii} \simeq D_{k,ii}$ the cross term reduces to $\ln L_{{\rm cross},k} \simeq -\tfrac12\sum_{i \ne j}(E_{k,ij} - S_{k,ij})^2/(D_{k,ii}D_{k,jj})$ up to a constant, a Gaussian in the predicted cross-powers centered on the measured cross-correlations and weighted by the auto-powers, whose maximum over $\Agw$ is the optimal statistic~\cite{Anholm2009,Chamberlin2015}.
In the present data the cross-power of every pulsar pair is a small fraction of the two auto-powers, so $\ln L_{\rm cross}$ is a much weaker function of $\Agw^2$ than $\ln L^i_{\rm auto}$ is of $A_{{\rm tot},i}^2$, which is the assumption made in the main text. In an array where the GWB dominated the variance of many pulsars this expansion would fail: the cross-correlations would then be constrained as well as the auto-correlations, and the volume effect would fade.

\emph{Sum over bins and ridge limit.}---Different frequency bins are independent, so we have $\ln L = \sum_k \ln L_k$. 
Let $\Nf$ be the number of modes in which a pulsar's red power exceeds its white noise. Its auto-correlation then fixes $A_{{\rm tot},i}^2$ to a relative width $\sim \Nf^{-1/2}$, confining the fit to a tube of that width around the ridge $A_{{\rm rn},i}^2 = A_{{\rm tot},i}^2 - \Agw^2$. As $\Nf \to \infty$ the tube collapses: the auto-correlations pin $A_{{\rm rn},i}^2 = A_{{\rm tot},i}^2 - \Agw^2$, and marginalizing each RN power against this sharply peaked gives,
\begin{equation}
\begin{aligned}
    p(\Agw^2|d) \propto\; &\pi_{\rm gwb}(\Agw^2)\, L_{\rm cross}(\Agw^2) \\
    &\times \prod_{i} \pi_{A^2}(A_{{\rm tot},i}^2 - \Agw^2)\,,
\end{aligned}
\end{equation}
valid for $0 \le \Agw^2 \le \min_i A_{{\rm tot},i}^2$, which is the many-mode limit $P^i_{\rm auto} \to \pi_{A^2}(A_{{\rm tot},i}^2 - \Agw^2)$ of Eq.~(\ref{eq:ridge}). Substituting the RN prior into the above equation and we can easily see the volume effect from different priors.

\begin{figure}[!t]
    \centering
    \includegraphics[width=\columnwidth]{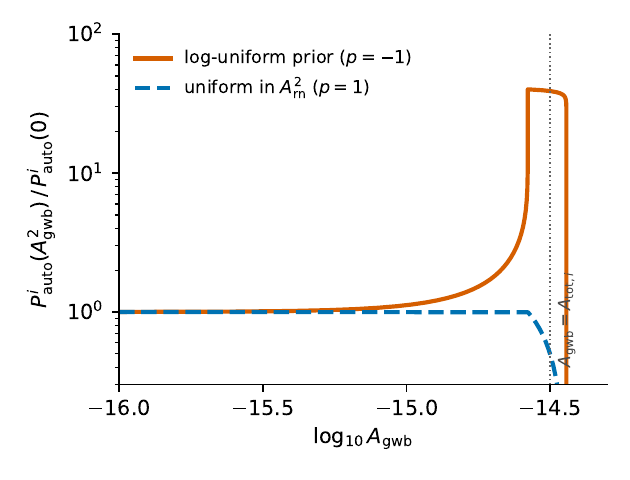}
    \caption{The per-pulsar auto-correlation factor $P^i_{\rm auto}(\Agw^2)$ of Eqs.~(\ref{eq:Pi_general}) and (\ref{eq:Pi_cases}) for one pulsar, normalized to its value as $\Agw \to 0$, under a prior uniform in the RN power ($p=1$, dashed) and the standard log-uniform prior ($p=-1$, solid). The pulsar's total red power is $A_{{\rm tot},i} = 10^{-14.5}$, measured to $\sigma_i = 0.3\,A_{{\rm tot},i}^2$, and the RN prior floor is $A_{{\rm rn},\min} = 10^{-20}$.}
    \label{fig:pauto}
\end{figure}

\emph{Finite width.}---At finite $\Nf$ the auto-correlation of pulsar $i$ fixes its total red power only to within $\sigma_i \sim A_{{\rm tot},i}^2\,\Nf^{-1/2}$. For a noisy pulsar $\Nf$ can be small, but the prior volume effect shows up similarly. To see this, let us approximate $L^i_{\rm auto}$ by a top hat of half-width $\sigma_i$ about the measured $A_{{\rm tot},i}^2$ (a Gaussian changes nothing qualitative), restricting the RN power in Eq.~(\ref{eq:ridge}) to the interval $[A_{{\rm tot},i}^2 - \sigma_i - \Agw^2,\, A_{{\rm tot},i}^2 + \sigma_i - \Agw^2]$, cut off below at the RN prior floor $A_{{\rm rn},\min}^2$. As $\Agw^2$ grows the interval slides down and the floor passes from below it to inside it to above it, which are the three cases of Eq.~(\ref{eq:Pi_cases}).
For $p \ne -1$ the per-pulsar factor (defined in Eq.~\ref{eq:ridge}) is then
\begin{equation}\label{eq:Pi_general}
\begin{aligned}
    P^i_{\rm auto}(\Agw^2) \propto\; &(A_{{\rm tot},i}^2 + \sigma_i - \Agw^2)^{\frac{p+1}{2}} \\
    &- [\max(A_{{\rm tot},i}^2 - \sigma_i - \Agw^2, A_{{\rm rn},\min}^2)]^{\frac{p+1}{2}}\,,
\end{aligned}
\end{equation}
where $A_{{\rm rn},\min}$ is the lower bound of the $A_{\rm rn}$ prior. For $p = 1$ (prior uniform in power space) the factor is constant for $\Agw^2 < A_{{\rm tot},i}^2 - \sigma_i$ and falls linearly to zero at $A_{{\rm tot},i}^2 + \sigma_i$: the auto-power sets only an upper limit on the GWB, and no $\Agw$ preference is introduced by the RN prior. For the log-uniform prior $p = -1$, writing $a_\pm \equiv A_{{\rm tot},i}^2 \pm \sigma_i - \Agw^2$ for the ends of that interval,
\begin{equation}\label{eq:Pi_cases}
    P^i_{\rm auto}(\Agw^2) =
    \begin{cases}
        \ln(a_+/a_-) & A_{{\rm rn},\min}^2 < a_-\,,\\[4pt]
        \ln(a_+/A_{{\rm rn},\min}^2) & a_- \le A_{{\rm rn},\min}^2 < a_+\,,\\[4pt]
        0 & A_{{\rm rn},\min}^2 \ge a_+\,.
    \end{cases}
\end{equation}
As $\Agw^2 \to 0$, $P^i_{\rm auto}$ approaches a constant $\ln[(A_{{\rm tot},i}^2+\sigma_i)/(A_{{\rm tot},i}^2-\sigma_i)] \simeq 2\sigma_i/A_{{\rm tot},i}^2$, independent of the RN floor. As $\Agw^2$ increases to about $A_{{\rm tot},i}^2 - \sigma_i$, $P^i_{\rm auto}$ rises to a plateau of height $\ln(2\sigma_i/A_{{\rm rn},\min}^2)$ and width $2\sigma_i$ in $\Agw^2$, where the GWB accounts for the pulsar's whole measured power and the RN may occupy every decade down to the prior floor. Finally $P^i_{\rm auto}$ drops to zero as $\Agw^2$ further increases. These three scenarios correspond to the left, middle, and right of the bottom panel of Fig.~\ref{fig:decomposition}. The difference is that the bottom of Fig.~\ref{fig:decomposition} shows the volume contribution so it increases monotonically with $\Agw^2$. $P^i_{\rm auto}$ here shows the contribution from the auto-correlation which includes both the profile likelihood and the volume term, so it approaches 0 in the high $\Agw$ limit. Figure~\ref{fig:pauto} shows a visualization of $P^i_{\rm auto}$ for $p=\pm1$.

\section{Per-pulsar volume breakdown}\label{app:volume}

The analytical model of Eq.~(\ref{eq:Pi_cases}) and Fig.~\ref{fig:pauto} predicts that under the log-uniform prior, pulsars with well measured red power (small $\sigma_i$) and the total red power not much larger than the GWB ($A_{{\rm tot}, i}\gtrsim \Agw$) contribute most to the volume effect. Here we test this on the real data, using the factorization of the volume term (Eq.~\ref{eq:volume_factorize}) which allows us to estimate the prior volume contribution from each individual pulsar. 

\begin{figure}[!t]
    \centering
    \includegraphics[width=\columnwidth]{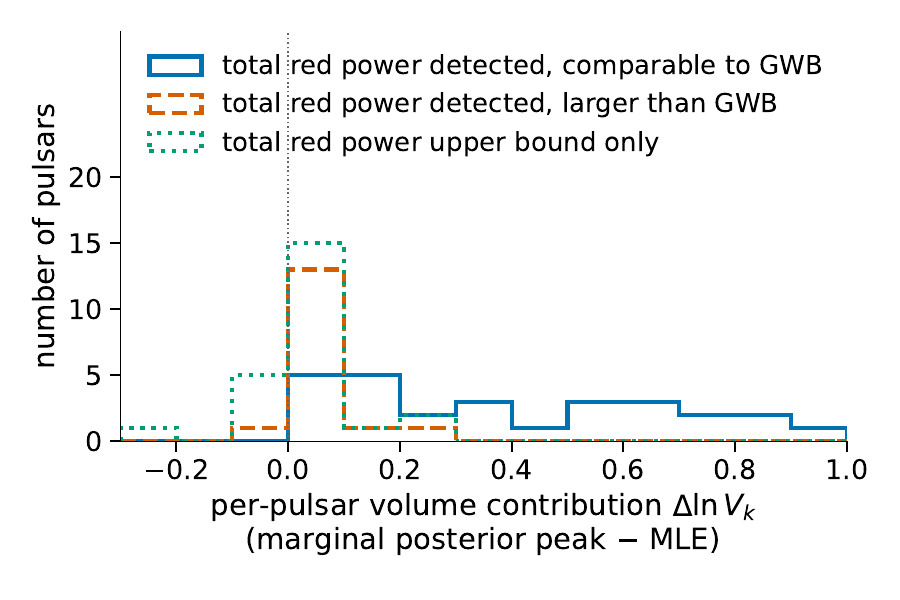}
    \caption{Per-pulsar contributions $\Delta\lnV_k$ to the prior volume term between the marginal-posterior peak, $(\logA_{\rm gwb},\ggw)=(-14.2,3.3)$, and the MLE, $(-14.5,3.6)$. }
    \label{fig:volume}
\end{figure}

Figure~\ref{fig:volume} shows the distribution of the contributions $\Delta\lnV_k$ between the marginal-posterior peak and the MLE, with the pulsars grouped by their total red power: whether it is detected, and whether it is comparable to the GWB. Comparing the MLE with the marginal posterior peak, the profile likelihood falls by $6.7$ while the volume term rises by $12.1$. The 27 pulsars with detected total red power comparable to the GWB supply $+11.0$ of the total $\Delta\lnV$, led by J1741+1351 ($+0.96$), J1738+0333 ($+0.85$), J1744$-$1134 ($+0.84$), J0023+0923 ($+0.75$), J2234+0611 ($+0.75$), J1713+0747 ($+0.69$) and J2229+2643 ($+0.66$). The 16 pulsars whose red power is much larger than the GWB contribute only $+0.6$ in total, and the 24 pulsars with only an upper bound on their total red power contribute $+0.5$.

The pulsars that dominate the volume are among the best-timed ones, with well measured total red power comparable to the GWB. This is expected, since only in these pulsars does the allowed RN range depend on $\Agw$: once the GWB accounts for their red power, the RN is released from a narrow interval around the measured power to the entire low-amplitude end of the prior, which under the log-uniform prior carries a large volume and inflates $\Agw$. Pulsars with much stronger intrinsic RN never release it, and pulsars with only an upper bound have it released already at the MLE, so neither group responds to $\Agw$. The volume effect is a consequence of adopting the log-uniform prior, not a defect of these good-quality pulsars.

\clearpage
\setcounter{section}{0}
\renewcommand{\thesection}{S\Roman{section}}
\setcounter{equation}{0}
\renewcommand{\theequation}{S\arabic{equation}}
\setcounter{figure}{0}
\renewcommand{\thefigure}{S\arabic{figure}}
\setcounter{table}{0}
\renewcommand{\thetable}{S\arabic{table}}
\onecolumngrid
\begin{center}
    {\large\bfseries Supplemental Material for ``Prior volume effect inflates the gravitational-wave background in pulsar timing arrays''}\\[6pt]
    Biwei Dai, Matias Zaldarriaga, and Yacine Ali-Ha\"{\i}moud
\end{center}
\twocolumngrid

\section{Consistency of the empirical Bayes prior}\label{app:eb}

We show that the empirical Bayes (EB) population prior converges to the true per-pulsar red-noise distribution as the number of pulsars grows, $\Np\to\infty$.

\emph{Setup.}---Suppose each pulsar's RN parameters $\bm\phi_k$ are drawn independently from a common population distribution $p_\star(\bm\phi)$, and that at fixed GWB parameters $\bm\theta$ the data of different pulsars are independent and follow per-pulsar likelihoods $\mathcal{L}_k(d_k|\bm\phi_k)$, which may differ from pulsar to pulsar through baseline, cadence, and white noise. This is exact for CURN. For HD the likelihood factorizes to leading order in the weak-cross-correlation expansion (see Eq.~\ref{eq:toy_factorize} in the End Matter), the cross term depending on the RN parameters only through the auto-power weights, and we work at that order, which is also the one at which the EM implementation below reweights per-pulsar marginals. We suppress the dependence on $\bm\theta$. For a candidate population prior $\pi$, the ensemble marginal likelihood factorizes,
\begin{equation}
\mathcal{M}(\pi) = \prod_{k=1}^{\Np} m_k(\pi), \qquad m_k(\pi) \equiv \int \mathcal{L}_k(d_k|\bm\phi)\,\pi(\bm\phi)\,d\bm\phi .
\end{equation}
The EB estimate is the nonparametric maximizer
\begin{equation}\label{eq:npmle}
\hat\pi = \arg\max_{\pi}\; \frac{1}{\Np}\sum_{k=1}^{\Np} \ln m_k(\pi),
\end{equation}
of which the iterative histogram reweighting used in the main text is a practical implementation (details in the Numerical Methods section below).

\emph{Limit of the objective.}---Write $m_k(\pi)=m_k(d_k;\pi)$ with $m_k(d;\pi)=\int \mathcal L_k(d|\bm\phi)\pi(\bm\phi)\,d\bm\phi$ the marginal density of pulsar $k$'s data under $\pi$. Since the $\bm\phi_k\sim p_\star$ are i.i.d.\ and $d_k\sim\mathcal L_k(\cdot|\bm\phi_k)$, the strong law of large numbers gives, for each fixed $\pi$,
\begin{equation}
\frac{1}{\Np}\sum_k \ln m_k(d_k;\pi) \;\xrightarrow{\ \Np\to\infty\ }\; Q(\pi) \equiv \overline{\mathbb{E}}\big[\ln m(d;\pi)\big],
\end{equation}
where $\overline{\mathbb E}$ averages over the population-and-kernel ensemble, with data generated from the true marginal $m(d;p_\star)$. Therefore
\begin{equation}
Q(\pi) - Q(p_\star) = -\,\overline{D}_{\rm KL}\!\big(m(\cdot;p_\star)\,\big\|\,m(\cdot;\pi)\big) \le 0,
\end{equation}
the (ensemble-averaged) Kullback--Leibler divergence, which is non-positive and vanishes if and only if $m(d;\pi)=m(d;p_\star)$ almost everywhere. Thus the population maximizer of the EB objective is any $\pi$ whose data \emph{marginal} matches the truth.

\emph{Identifiability.}---If the mixture map $\pi\mapsto m(\cdot;\pi)$ is injective, i.e.\ distinct population priors produce distinct data marginals, then $m(\cdot;\pi)=m(\cdot;p_\star)$ implies $\pi=p_\star$. The power-law RN model satisfies this requirement~\cite{Teicher1961,Teicher1963}: each pulsar's data resolve the red-noise power at many Fourier frequencies, so no two distinct population distributions over $(\gamma,\logA)$ reproduce the same data marginal. Standard M-estimation (Wald) consistency then yields
\begin{equation}
\hat\pi \;\xrightarrow{\ \Np\to\infty\ }\; p_\star
\end{equation}
(weak convergence): the EB prior recovers the true population distribution in the many-pulsar limit. This is the Kiefer--Wolfowitz consistency~\cite{KieferWolfowitz1956,Lindsay1995} of the nonparametric maximum-likelihood estimate of a mixing distribution, here with heterogeneous per-pulsar kernels.

\emph{EM iteration.}---The maximizer $\hat\pi$ is found by expectation maximization~\cite{Laird1978}. Given the current estimate $\pi_t$, the E-step computes each pulsar's posterior $p(\bm\phi|d_k,\pi_t)\propto\mathcal L_k(d_k|\bm\phi)\,\pi_t(\bm\phi)$, and the M-step replaces the population estimate by their average, $\pi_{t+1}(\bm\phi)=\Np^{-1}\sum_k p(\bm\phi|d_k,\pi_t)$. Each step increases the objective of Eq.~(\ref{eq:npmle}), which is concave in $\pi$ as a sum of logarithms of linear functionals, so every stationary point is a global maximum and the iteration converges to $\hat\pi$ rather than to a local optimum. Combined with the consistency result, the EM iteration therefore converges to the true population distribution as $\Np\to\infty$. In practice the E-step is carried out by importance reweighting of the samples of the standard HD chain and the density is represented as a smoothed histogram (see Numerical methods below).

\emph{Remarks.}---(i)~The result is asymptotic in $\Np$; at finite $\Np$ the estimate fluctuates about $p_\star$ at the $\mathcal{O}(\Np^{-1/2})$ level. (ii)~Because $\hat\pi\to p_\star$, the prior volume bias vanishes in the many-pulsar limit \emph{for the EB (or true population) prior}, whereas a fixed, population-independent prior such as the log-uniform retains an $\mathcal{O}(\Np)$ bias [Eq.~(\ref{eq:ridge})]. (iii)~The argument holds at fixed GWB parameters $\bm\theta$, or under joint estimation, with the per-pulsar evidences evaluated in the conditional likelihood $\mathcal L_k(d_k|\bm\phi,\bm\theta)$, and for HD to leading order in the cross-correlations. Marginalizing over $\bm\theta$ does not affect the limit: $\bm\theta$ is a finite-dimensional parameter common to all pulsars, and the joint estimate of such a parameter and of the population distribution is consistent~\cite{KieferWolfowitz1956}. 

\section{Evidence for Hellings-Downs correlations}\label{app:bfhdcurn}

Here we explain why the revised priors raise the Bayes factor between the HD and CURN models (Table~\ref{tab:bf_hdcurn}) even though they lower the inferred amplitude.

\emph{Amplitude-resolved Bayes factor.}---Write $a\equiv\logA_{\rm gwb}$ and let $\bm\phi$ denote the red-noise parameters. For each model $M\in\{{\rm HD},{\rm CURN}\}$ define the evidence at fixed amplitude $Z_M(a)$, the total evidence $Z_M$, the marginal amplitude posterior $p_M(a)$, and the Bayes factor at fixed amplitude $\mathrm{BF}(a)$,
\begin{equation}\label{eq:zcond}
\begin{aligned}
    Z_M(a) &\equiv \int\! d\bm\phi\; \mathcal{L}_M(d|a,\bm\phi)\,\pi(\bm\phi), \\
    Z_M &\equiv \int\! da\; \pi(a)\,Z_M(a), \\
    p_M(a) &\equiv \pi(a)\,Z_M(a)/Z_M, \\
    \mathrm{BF}(a) &\equiv Z_{\rm HD}(a)/Z_{\rm CURN}(a),
\end{aligned}
\end{equation}
so that $\mathrm{BF}=Z_{\rm HD}/Z_{\rm CURN}$. Substituting the definitions,
\begin{equation}\label{eq:bf_of_A}
    \int\! da\, \frac{p_{\rm HD}(a)}{\mathrm{BF}(a)}
      = \frac{\int\! da\, \pi(a)\,Z_{\rm CURN}(a)}{Z_{\rm HD}} = \frac{1}{\mathrm{BF}},
\end{equation}
we see that the Bayes factor is the harmonic mean of the fixed-amplitude Bayes factor over the amplitude posterior of the HD model (equivalently, its arithmetic mean over the CURN posterior). Although each evidence is a \emph{prior} average of the likelihood, their ratio is a \emph{posterior} average of the pointwise ratio, which is why the amplitude at which the HD posterior sits controls the Bayes factor: amplitudes carrying no posterior weight cannot contribute, however much prior mass they hold.

\begin{figure}[t]
    \centering
    \includegraphics[width=\columnwidth]{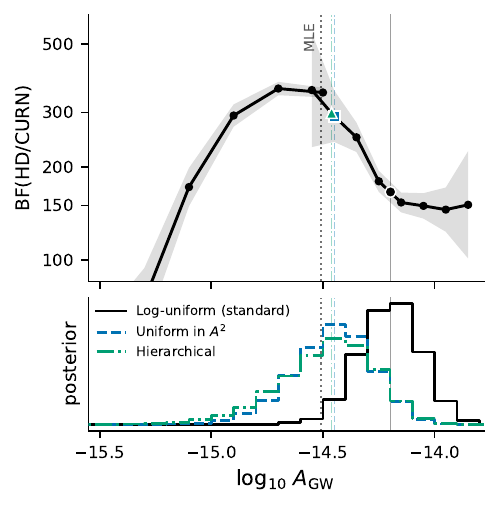}
    \caption{Amplitude-resolved Bayes factor. \textbf{Top:} $\mathrm{BF}(\Agw)$, the HD-versus-CURN Bayes factor at fixed GWB amplitude. The band is $1\sigma$ from the effective number of samples per bin. \textbf{Bottom:} the HD-model posterior on $\Agw$ under each of the three priors, which is the weight Eq.~(\ref{eq:bf_of_A}) applies to the curve above. Vertical lines mark the posterior medians, and the dotted line mark the MLE. The standard log-uniform prior displaces the posterior weight to an amplitude where the fixed-amplitude Bayes factor is about half its value near the MLE, which is how the prior volume effect suppresses the evidence for Hellings-Downs correlations.}
    \label{fig:bf_vs_a}
\end{figure}

Figure~\ref{fig:bf_vs_a} shows $\mathrm{BF}(a)$ together with the posterior weight that Eq.~(\ref{eq:bf_of_A}) averages it against. It is peaked, and the peak sits at the maximum-likelihood amplitude: $\mathrm{BF}(a)$ rises from ${\simeq}\,170$ at $\logA_{\rm gwb}=-15.1$ to ${\simeq}\,350$ at $-14.5$ and falls to ${\simeq}\,145$ at $-13.9$. It must fall on the low side, since the two models coincide as $\Agw\to0$ and the Bayes factor tends to unity; it falls on the high side because the HD correlations measure the cross-power, on which the per-pulsar RN priors have no volume to spend, so displacing $\Agw$ above the amplitude the cross-correlations support costs HD likelihood that CURN does not pay. The standard HD posterior sits well above the peak, at $\logA_{\rm gwb}=-14.20$, where the fixed-amplitude Bayes factor is ${\simeq}\,165$, whereas the two mitigated posteriors sit at $-14.45$ and $-14.46$, on the peak, where it is ${\simeq}\,300$. 

A single curve serves all three runs because $\mathrm{BF}(a)$ is nearly prior independent, as expected since the red-noise prior largely cancels between $Z_{\rm HD}(a)$ and $Z_{\rm CURN}(a)$. Read at each prior's own HD posterior median the curve returns $166$, $291$ and $299$ against measured totals of $176$, $263$ and $300$: to ${\sim}\,15\%$ the three Bayes factors are one curve sampled at three amplitudes.

\section{Numerical methods}\label{app:methods}

\emph{Profile likelihood.}---The profile likelihood is measured with differential evolution with 192 points followed by L-BFGS-B refinement with a finite-difference gradient. Grid points are warm-started from converged neighbors, and a reconciliation pass re-optimizes any point lying below a neighbor. The global maximum was confirmed by 5 independent DE runs of different seeds.

\emph{Profile-likelihood confidence levels.}---We convert the profile likelihood to confidence regions with Wilks' theorem~\cite{Wilks1938}: asymptotically, $2\Delta\ln\mathcal{L}_{\rm prof}$, the drop below the maximum, follows a $\chi^2$ distribution whose number of degrees of freedom equals the number of parameters not profiled over. For the 1D profile in $\logA_{\rm gwb}$ at fixed $\ggw=13/3$ (Fig.~\ref{fig:fixgamma}), the $90\%$ interval is $2\Delta\ln\mathcal{L}_{\rm prof}\le2.706$. For the 2D profile in $(\ggw,\logA_{\rm gwb})$ (Fig.~\ref{fig:decomposition}), the $1\sigma$ and $2\sigma$ contours enclose the Gaussian coverages $p=68.3\%$ and $95.4\%$, which for two degrees of freedom correspond to $\Delta\ln\mathcal{L}_{\rm prof}=-\ln(1-p)=1.15$ and $3.09$.

\emph{Volume integration.}---Each per-pulsar factor $V_k(\bm\theta)$ (Eq.~\ref{eq:Vi}) is a 2D integral over $(\gamma_k,\logA_k)$ evaluated numerically on a grid. Numerical convergence was verified. We do not use the Laplace approximation of Ref.~\cite{Hadzhiyska2023}, because the per-pulsar RN is poorly constrained and its likelihood is far from Gaussian (bottom row of Fig.~\ref{fig:decomposition}).

\emph{Empirical Bayes prior.}---The EM iteration is run on the per-pulsar $(\gamma_k,\logA_k)$ samples of the standard log-uniform HD chain. Because that reference prior is flat in $(\gamma,\logA)$, the importance weight of a sample under the current population estimate $p_t$ is simply $p_t$ evaluated at that sample. Each pulsar's weighted samples are normalized separately, so every pulsar contributes equally to the EM average irrespective of its sample count. The density is represented as a $50\times50$ histogram on $[0,7]\times[-20,-11]$, smoothed after each step by an elliptical Gaussian kernel of width $1.5$ bins whose orientation tracks the running population correlation, and iterated $100$ times from a uniform start. The smoothing regularizes the well-known tendency of the nonparametric maximum-likelihood estimate to collapse onto discrete atoms at finite $\Np$.

\emph{MCMC.}---Posteriors are sampled with \texttt{PTMCMCSampler}~\cite{PTMCMCSampler}, on likelihoods built with \texttt{enterprise}~\cite{enterprise} and \texttt{enterprise\_extensions}~\cite{enterprise_extensions}, with
$5\times10^6$ iterations, and white noise fixed to the MAP. The Bayes factors of Table~\ref{tab:bf_hdcurn} are computed by sampling HD and CURN jointly in a product space with a discrete model index~\cite{CarlinChib1995,Hee2016} and a constant log-likelihood weight on CURN to balance the occupancies~\cite{Johnson2024}; the quoted uncertainties follow from the effective sample size of the model indicator, and the standard-prior value is consistent with the published $226\pm70$~\cite{NANOGrav15_gwb}.

\emph{External comparison constraints.}---%
The PPTA~\cite{PPTA_gwb} reports a $68\%$ interval, which we convert to $90\%$ by scaling its half-width by $1.645$, assuming a Gaussian posterior.

\emph{Astrophysical predictions.}---The two astrophysical intervals in Fig.~\ref{fig:fixgamma} are built from the predictions of Ref.~\cite{Sato-Polito2024} for the characteristic strain at $f=\mathrm{yr}^{-1}$ with $\ggw=13/3$, each obtained by combining a galaxy distribution function with a black hole scaling relation. The stellar-mass row combines one galaxy stellar mass function with two calibrations of the $M_{\rm BH}$--$M_{\rm bulge}$ relation, and the velocity-dispersion row combines three velocity dispersion functions with two calibrations of the $M_{\rm BH}$--$\sigma$ relation, all as listed in Ref.~\cite{Sato-Polito2024}. Within each row we take the mean of the individual predictions as the central value, add the statistical uncertainty and the spread between combinations in quadrature, and convert the resulting $68\%$ interval to $90\%$ assuming a log-normal distribution. This gives $A_{\rm gwb}=1.36^{+0.63}_{-0.43}\times10^{-15}$ for the stellar-mass route and $1.04^{+0.51}_{-0.34}\times10^{-15}$ for the velocity-dispersion route.

\emph{Decoupled HD model and its positive-definiteness boundary.}---The model sets the diagonal of the HD overlap-reduction function to zero, $\Gamma_{\rm HD}(\theta_{aa})\to0$, so the HD component contributes only to inter-pulsar cross-correlations. Two distinct restrictions on $(A_{\rm auto}, A_{\rm cross})$ must then be separated. Writing Eq.~(\ref{eq:cross_only}) as $C(f) = S_{\rm auto}(f)\,\mathbb{I} + S_{\rm cross}(f)\,\tilde\Gamma$ with $\tilde\Gamma \equiv M - \tfrac12\mathbb{I}$, where $M_{ab}=\Gamma_{\rm HD}(\theta_{ab})$ is the full earth-term overlap matrix (including $M_{aa}=\tfrac12$) and is positive semidefinite, the eigenvalues of $\tilde\Gamma$ are bounded below by $-\tfrac12$. Positive definiteness of $C$ therefore requires only
\begin{equation}\label{eq:psd}
    S_{\rm auto}(f) \;\ge\; |\lambda_{\min}|\,S_{\rm cross}(f), \qquad |\lambda_{\min}| \le \tfrac12 ,
\end{equation}
and for the 67-pulsar array $\lambda_{\min}=-0.4995$. This is the condition we impose. It is strictly weaker, by a factor of two in power, than the condition $S_{\rm auto}\ge S_{\rm cross}$ that would follow from requiring the two components to be a physical CURN-plus-GWB pair with non-negative amplitudes, and we deliberately do not impose the latter. The reason is that $S_{\rm auto}=S_{\rm cross}$ is exactly the location of zero CURN, so for a pure GWB the truth sits \emph{on} that boundary. Imposing it would make the inferred $A_{\rm cross}$ depend on the measured auto-power, reintroducing the prior dependence in the auto-power. What the model estimates is the amplitude of the HD correlation pattern, a nuisance-projected quantity, and truncating a noisy estimator at a physical boundary is itself a source of bias. Empirically, $30\%$ of the real-data posterior lies at $S_{\rm cross}>S_{\rm auto}$. Restricting to the physical region would move $\logA_{\rm cross}$ from $-14.69$ to $-14.73$ and its $90\%$ upper edge from $-14.52$ to $-14.58$, slightly \emph{away} from the standard-prior value and leaving our conclusion intact. In the phase-shifted null the boundary is entirely inactive so the quoted upper limit is unaffected.

This is also why both slopes are held at $13/3$. With $\gamma_{\rm auto}=\gamma_{\rm cross}$ the ratio $S_{\rm auto}/S_{\rm cross}$ is frequency independent and Eq.~(\ref{eq:psd}) collapses to a single inequality on the amplitude ratio. If the slopes differ the condition must hold in \emph{every} frequency bin, so the effective constraint is set by the worst one and is strictly tighter, leading to more leakage from the auto-power, and because the cross channel prefers a steeper spectrum than the auto channel that bin is normally the lowest frequency. 

\section{New-physics Bayes factors for all source models}\label{app:newphysics_full}

Table~\ref{tab:newphysics_full} lists the Bayes factors $\mathrm{BF}(\mathrm{model}/\mathrm{SMBHB})$ for all fifteen new-physics source models of the NANOGrav 15-year search~\cite{NANOGrav15_newphysics} under each per-pulsar RN prior, of which Table~\ref{tab:newphysics} shows the highest-BF variant of each source class. Under the standard log-uniform prior our Bayes factors are consistent with those of Ref.~\cite{NANOGrav15_newphysics}, validating the factorized refit.

\begin{table*}[t]
    \centering
    \caption{Bayes factors $\mathrm{BF}(\mathrm{model}/\mathrm{SMBHB})$ for all fifteen new-physics GWB source models of the NANOGrav 15-year new-physics search~\cite{NANOGrav15_newphysics}, under the standard log-uniform, uniform-in-$A^2$, hierarchical Gaussian, and empirical Bayes (EB) per-pulsar RN priors. Model keys follow the \texttt{PTArcade} model library~\cite{PTArcade}.}
    \label{tab:newphysics_full}
    \begin{tabular}{llcccc}
    \hline
    Source & Model & Standard & Uniform $A^2$ & Hier. & EB \\
    \hline
    Scalar-induced GW  & \texttt{sigw\_gauss}  & $57\pm16$ & $0.50\pm0.11$ & $2.3\pm0.5$ & $4.6\pm1.2$ \\
                       & \texttt{sigw\_delta}  & $38\pm10$ & $0.56\pm0.13$ & $3.2\pm0.8$ & $5.9\pm1.6$ \\
                       & \texttt{sigw\_box}    & $22\pm8$ & $0.27\pm0.06$ & $0.83\pm0.22$ & $1.8\pm0.5$ \\
    Superstrings       & \texttt{super}        & $32\pm8$ & $0.92\pm0.18$ & $4.2\pm0.9$ & $7.0\pm1.8$ \\
    Metastable strings & \texttt{meta\_ls}     & $23\pm7$ & $0.79\pm0.16$ & $1.8\pm0.4$ & $3.1\pm0.8$ \\
                       & \texttt{meta\_l}      & $15\pm4$ & $0.74\pm0.16$ & $1.6\pm0.4$ & $2.7\pm0.7$ \\
    Domain walls       & \texttt{dw\_sm}       & $19\pm6$ & $0.36\pm0.09$ & $0.79\pm0.20$ & $1.4\pm0.4$ \\
                       & \texttt{dw\_ds}       & $1.8\pm0.5$ & $0.41\pm0.11$ & $0.69\pm0.17$ & $1.2\pm0.3$ \\
    Phase transition   & \texttt{pt\_bubble}   & $15\pm4$ & $0.31\pm0.07$ & $1.0\pm0.2$ & $1.7\pm0.4$ \\
                       & \texttt{pt\_sound}    & $5.8\pm2.2$ & $0.23\pm0.06$ & $0.42\pm0.11$ & $0.66\pm0.18$ \\
    Inflation          & \texttt{igw}          & $6.4\pm2.7$ & $0.06\pm0.02$ & $0.27\pm0.08$ & $0.53\pm0.15$ \\
    Stable strings     & \texttt{stable\_m}    & $0.21\pm0.06$ & $1.7\pm0.3$ & $1.7\pm0.4$ & $1.7\pm0.4$ \\
                       & \texttt{stable\_k}    & $0.24\pm0.06$ & $1.5\pm0.3$ & $1.7\pm0.4$ & $1.5\pm0.3$ \\
                       & \texttt{stable\_n}    & $0.23\pm0.06$ & $1.5\pm0.3$ & $1.6\pm0.3$ & $1.3\pm0.3$ \\
                       & \texttt{stable\_c}    & $0.20\pm0.05$ & $1.5\pm0.3$ & $1.5\pm0.3$ & $1.2\pm0.3$ \\
    \hline
    \end{tabular}
\end{table*}

\end{document}